# Geometric and Dosimetric Validation of Deformable Image Registration for Prostate MR-guided Adaptive Radiotherapy

Victor N. Malkov [0,1,2], Iymad R. Mansour[1,2], Vickie Kong [1,2], Winnie Li [1], Jennifer Dang[1], Parisa Sadeghi [1], Inmaculada Navarro[1], Jerusha Padayachee [1,2], Peter Chung [1,2], Jeff D. Winter[1,2], Tony Tadic [1,2]

[0]Mayo Clinic, Department of Radiation Oncology, Rochester, MN, USA

[1]Radiation Medicine Program, Princess Margaret Cancer Centre, Toronto, Ontario, Canada,

[2]Department of Radiation Oncology, University of Toronto, Toronto, Ontario, Canada

*This is the version of the article before peer review or editing at the time of submission by the authors.*

Corresponding Author address: Malkov.Victor@mayo.edu

Funding: None

Conflicts of Interest: None

Data Availability Statement: No, my research data will not be publicly available

Short running title**:** geometric and dosimetric DIR validation for prostate MR ART

## Abstract

**Objective:** To quantify geometric and dosimetric accuracy of a novel prostate MR-to-MR deformable image registration (DIR) approach to support MR-guided adaptive radiation therapy dose accumulation.

**Approach**: We evaluated DIR accuracy in 25 patients treated with 30 Gy in 5 fractions on a 1.5 T MR-linac using an adaptive workflow. For all patients, a reference MR was used for planning, with three images collected at each fraction: adapt MR for adaptive planning, verify MR for pretreatment position verification and beam-on for capturing anatomy during radiation delivery. We assessed three DIR approaches: intensity-based, intensity-based with controlling structures (CS) and novel intensity based with controlling structures and points of interest (CS+P). DIRs were performed between the reference and fraction images and within fractions (adapt-to-verify and adapt-to-beam-on). For the evaluation we propagated CTV, bladder, and rectum contours using the DIRs and compared each to manually delineated contours using Dice similarity coefficient, mean distance to agreement, and dose-volume metrics.

**Results**: CS and CS+P improved geometric agreement between manual and propagated contours over intensity-only DIR. For example, mean distance to agreement ($DTA_{mean}$) for reference-to-beam-on intensity-only DIR was 0.131±0.009 cm (CTV), 0.46±0.08 cm (bladder), and 0.154±0.013 cm (rectum). For the CS, the values were 0.018±0.002 cm, 0.388±0.14 cm, and 0.036±0.013 cm. Finally, for CS+P these values were 0.015±0.001 cm, 0.025±0.004 cm, and 0.021±0.002 cm. Dosimetrically, comparing CS and CS+P for reference to beam-on DIRs resulted in a change of CTV D98% from [-29 cGy, 19 cGy] to [-18 cGy, 26 cGy], rectum D1cc from [-106 cGy, 72 cGy] to [-52 cGy, 74 cGy], and bladder D5cc from [-51 cGy, 544 cGy] to [-79 cGy, 36 cGy].

**Significance:** CS improved geometric and dosimetric accuracy over intensity-only DIR, with CS+P providing the most consistent performance. However, session image segmentation remains a challenge, which may be addressed with automated contouring.

# Introduction

Integrated MRI linear accelerator (MRL) systems offer superb soft tissue contrast, real-time imaging during radiation delivery for targeting monitoring and gating, as well as support for online adaptive radiation therapy (ART) to tailor each treatment to the daily internal anatomy (Winkel et al., 2019). With a daily adaptive framework, cumulative dose delivered over the full treatment course includes inter-fractional anatomical differences, unique daily adaptive treatment plans and internal anatomical changes between adapted plan generation and beam delivery (Brennan et al., 2023; de Muinck Keizer et al., 2020). Quantifying accumulated delivered dose following each fraction enabling monitoring of aggregate target and OAR doses relative to the original reference plan and treatment clinical goals, offering an opportunity to make informed adaptive replanning decisions for future fractions. Moreover, accumulated dose can be leveraged to evaluate and optimize adaptive strategies (Fredén et al 2022, Murr et al 2024b), as well as provide a more meaningful correlate between dose volume metrics and patient outcomes (Bohoudi et al., 2021; Willigenburg et al., 2022). Using deformable image registration (DIR) it is possible to map dose to a target image set with the associated deformed vector field (DVF) enabling dose summation that can be leveraged for calculation of accumulated delivered dose on the MRL (Chetty & Rosu-Bubulac, 2019; Jaffray et al., 2010; McDonald et al., 2023; Nenoff et al., 2023). However, dose accumulation accuracy is directly related to DIR performance, as such dose accumulation tools require careful application-specific validation prior to clinical use.

Validation and quantification of uncertainties in the DIR used to deform dose for accumulation is a critical step required prior to clinical dose accumulation implementation in

MR-guided ART. The AAPM Task Group 132 and more recent reports describe quantitative and qualitive tools required for establishing the accuracy of deformable image registration (Brock et al., 2017; Nenoff et al., 2023). Validation requires expert visual review as well as geometric and dosimetric considerations. Large anatomical changes or gain/loss of image information between the target and moving images are particularly challenging for DIR performance. Both scenarios exist in the inter- and intra-fraction bladder and rectum changes within prostate MRL adaptive workflows necessitating novel DIR solutions (Lorenzo Polo et al., 2024). Recent work compared various DIR approaches currently employed at different centres for multiple treatments sites for the purpose of inter-fraction dose accumulation (Murr, Bernchou, et al., 2024). This cited study demonstrated generally high between-institution agreement with bladder dose-volume differences being the greatest for the prostate cases.

In this work, we proposed a novel solution to account for large deformations for both inter- and intra-fraction prostate MRL DIR using a hybrid image-intensity approach based on controlling structure and points. We performed a geometric and dosimetric evaluation of this novel approach including comparison with conventional intensity-only or hybrid intensity and controlling structure DIR approaches.

## Materials and Methods:

**Patients**

We investigated DIR accuracy in a cohort of 25 patients enrolled in a prospective clinical trial investigating MR-guided prostate brachytherapy and external beam radiation therapy (ClinicalTrials.gov Identifier: NCT00913939). The study was approved by our institution's research ethics board and all patients provided informed written consent. Patients received 30

Gy in 5 fractions treated on the MR-Linac with consecutive 15 Gy in 1 high dose brachytherapy boost to intraprostatic gross tumour volumes.

**MR-Linac Treatment**

We performed reference and online treatment planning using Monaco 5.40.01 (Elekta, Stockholm, Sweden) with a 9-beam IMRT beam geometry designed to meet our institutional dose constraints (Table S1). Monte Carlo dose was calculated on a 3 mm grid incorporating the 1.5 T magnetic field with 1 % statistical uncertainty. At each adaptive fraction, we acquired an MR ($MR_{adapt}$) and performed a manual rigid registration with the reference MR based on a soft tissue match of the prostate. A radiation oncologist or fellow performed any required manual contour edits based on rigid or deformably propagated contours, focusing on the region with 2 cm of the planning target volume. Next, we generated an adapted treatment plan based on small real-time adjustments of the starting objectives using the newly contoured structures. For dose computation, bulk densities were assigned to the external, femurs, and PTV using mean electron densities extracted from a CT image collected on the same day as the MR simulation. As part of our standard workflow, prior to treatment delivery, we acquired and reviewed a verification MR ($MR_{verify}$) collected during plan quality assurance, followed by a 6-minute beam-on MR ($MR_{beam-on}$) during beam delivery for potential offline evaluation. Table S2 of the supplement provides the imaging parameters for the MR sequences.

**Fractional dose computation and manual contouring**

For each fraction we recomputed the fractional adapted treatment plan on each of the $MR_{verify}$ and $MR_{beam-on}$ images using offline Monaco. We then imported the reference plan dose, reference MR, fractional doses and MR images ($MR_{adapt}$, $MR_{verify}$, $MR_{beam-on}$) into RayStation 10B

(RaySearch, Stockholm, Sweden). In total we analyzed for 25 reference and 375 fractional MR images structures and doses.

For offline contouring, a radiation-oncologist-trained adaptive radiation therapist reviewed and edited CTV (prostate), bladder and rectum on the $MR_{reference}$ and $MR_{adapt}$, and generated manual contours on the $MR_{verify}$ and $MR_{beam-on}$ in RayStation following the ESTRO ACROP recommendations (Salembier et al., 2018). Resulting contours were inspected and modified as required by an independent RO fellow and confirmed by a medical physicist who flagged outliers and prompted an additional contour edits and review process iterations.

**Controlling point generation**

An inhouse-developed script automatically generated anchor points on the surface of the bladder, rectum and CTV structures to be used as controlling points in the DIR generation. These points are meant to represent corresponding anatomical locations on the organ surfaces between the MR images. To determine these point locations, we generated mesh representations of the manually contoured structures on the reference and fractional MR images using the RayStation *create controlling ROI* function (mesh detail = 0.4). This function first generates a 3D mesh on each image that is a smooth approximation to the manual contours. Each mesh is based on a common topology using the same number of vertices and indexing of edges, ensuring a direct vertex-to-vertex mapping.

The set of controlling POIs for DIR generation were selected as a subset of the mesh vertices on the reference MR. Starting from a random mesh vertex, subsequent vertices were selected ensuring a minimum separation of 10 mm as calculated using the dijkstra_path_length function from the NetworkX python library (Hagberg et al., 2008). The selected mesh indices were then used to identify the corresponding coordinates for the controlling POIs on the fractional images.

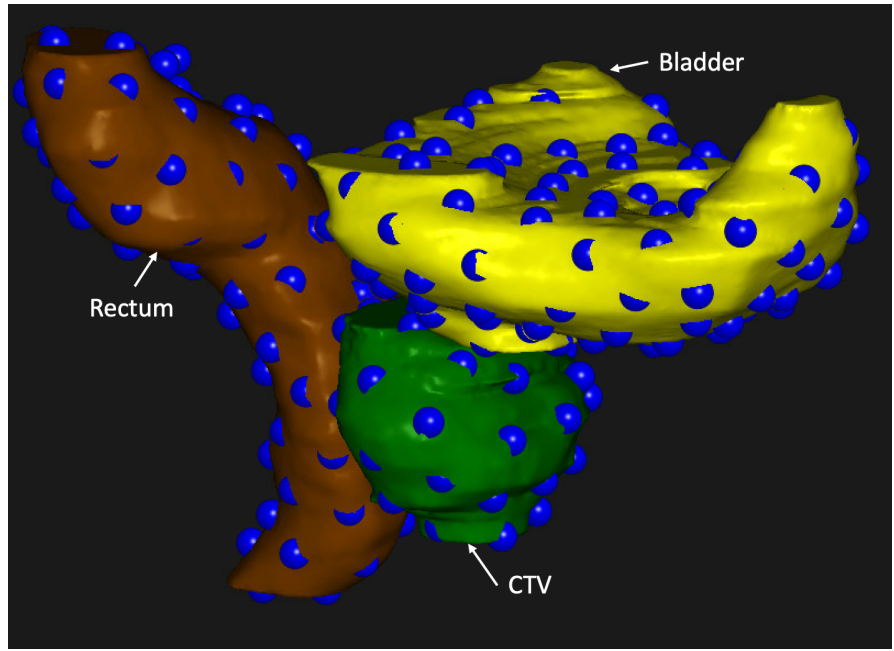

Figure 1: 3D visualization of controlling points (blue spheres) generated on the surface of the bladder (yellow), CTV (green), and rectum (brown) for a representative patient.

**Deformable image registration**

We generate DIRs in RayStation using the hybrid intensity- and structure-based algorithm called ANACONDA (Weistrand & Svensson, 2015a). Our analysis included interfraction DIRs, between the reference MR and each of the three fractional MR images, and intrafraction DIRs, between the daily planning MR ($MR_{adapt}$) and verification ($MR_{verify}$) and beam-on ($MR_{beam-on}$) images. We use the shorthand nomenclature: R2A to represent reference to adapt MR DIRs and similarly, R2V (reference to verify), R2B (reference to beam-on), A2V (adapt to verify) and A2B (adapt to beam-on). Figure 2a summarizes the DIR generation and nomenclature. We followed a stepwise study design to assess potential improvement in DIR accuracy using either: 1. Image-intensity only, 2. Image-intensity with controlling structures (CS), and 3. Image-intensity with controlling structures and points (CS+P). We utilized the manual CTV, bladder, and rectum contours as the controlling structures. Controlling points were defined for each controlling structure using the mesh representations via the method described above. Overall, for each image pair and DIR strategy 125 DIRs were performed (25 patients x 5 fractions).

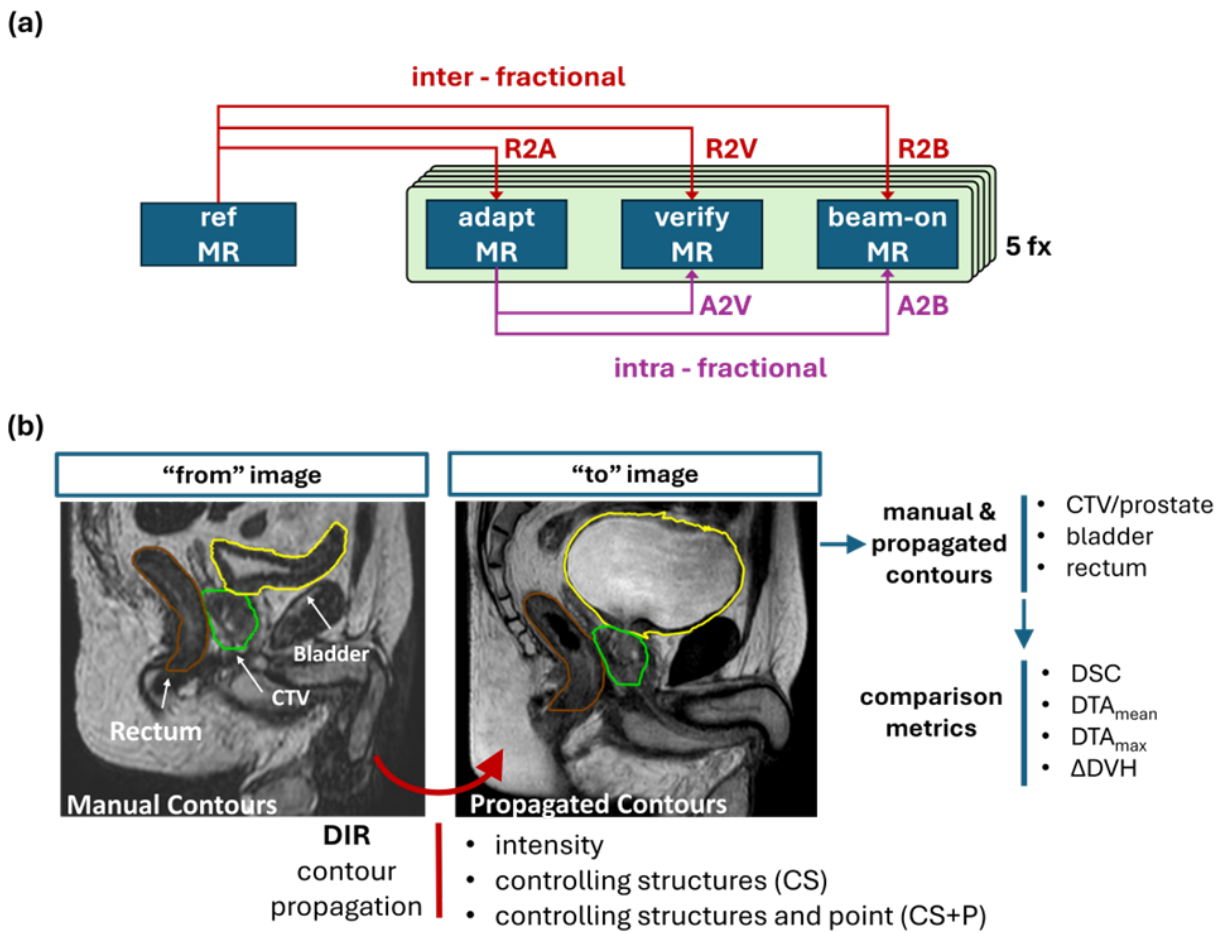

*Figure 2: (a) Summary of image pairs (inter or intra fractional) used in DIR generation and evaluation and (b) Overview of the DIR strategies used for mapping contours between image pairs and the comparison metrics used for DIR accuracy evaluation. Three DIR strategies (intensity-only, CS, CS+P) are used to deform the "from" image CTV, bladder, and rectum to the "to" image.*

*DIR evaluation is performed using geometry and DVH metrics comparing the propagated contours to the manual contours on the "to" image.*

**DIR Analysis**

To assess DIR accuracy, we applied the DIR strategies (intensity, CS, or CS+P) to deformably propagate the bladder, rectum, and CTV manually delineated contours between the "from" and "to" images. We then performed a geometric comparison of the propagated contours with the corresponding manual contours on the "to" image by computing the volume Dice Similarity Coefficient (DSC), mean distance to agreement ($DTA_{mean}$) and the max distance to agreement ($DTA_{max}$) using scripting methods in RayStation.

To further assess dosimetric DIR accuracy, we extracted DVHs for each of the propagated and manual contours using the reconstructed dose distributions on each image (no dose propagation is performed). We computed the difference of these relative volume DVHs, $\Delta$DVH, to provide a dose-informed evaluation of the DIR accuracy. In this way, the $\Delta$DVH represents the dosimetric error arising from geometric misalignment of the propagated and manual contours, ultimately due to a combination of manual contouring and DIR uncertainties. The $\Delta$DVHs were scaled to the full treatment course for comparison with reference plan clinical goals. We also report a clinically relevant set of specific DVH metrics from the $\Delta$DVH for each image pair and DIR approach: CTV D98%, bladder D5cc, and rectum D1cc. The DIR strategies and evaluation metrics are summarized in Figure 1b.

**Results**

The results below focus on the A2B and R2B DIR as these would be of greatest utility in the context of interfraction and intrafraction dose accumulation. Detailed results for the other DIRs are presented in the supplementary materials.

**Intra-fractional**

In figure 3, we provide the $DTA_{max}$, $DTA_{mean}$, and the DSC plots for the A2V and A2B image pairs for the intensity, CS, and CS+P DIR strategies. Table 1 provides the mean for each metric and image pair. The CS strategy produces an improvement of all metrics for the CTV and the rectum relative to the intensity only approach. For example, for the A2B pair, $DTA_{mean}$ changed from 0.085 ± 0.007 cm to 0.015 ± 0.002 cm for the CTV and 0.118 ± 0.008 cm to 0.024 ± 0.006 cm for rectum. Generally, the CS+P DIR performance did not substantially differ from the CS approach for structures with minimal geometric variations, however measurable improvements existed for structures exhibiting large geometric variations. For example, CS+P improved the DIR performance for bladder with $DTA_{mean}$ for A2B reducing from 0.47 ± 0.07 cm to 0.27 ± 0.12 cm between intensity and CS DIRs, with only a small subset of fractions performing worse than the intensity-only DIR approach. This is most pronounced in the A2B comparisons and is attributed to bladder filling over the course of the adaptive fraction and is presented in further detail in the inter-fraction results. In Figure 4, we provide the manual contours on the moving image and propagated contours using all three DIR strategies for a patient with a poor bladder deformation using the CS method. We observed that the reference image had a near empty bladder and results in poor propagation where the entire bladder is directed towards the inferior bladder surface. The CS+P strategy corrects these deviations as seen in the improved geometry metrics for the bladder ($DTA_{mean}$ reducing to 0.022 ± 0.003 cm for the A2B mapping).

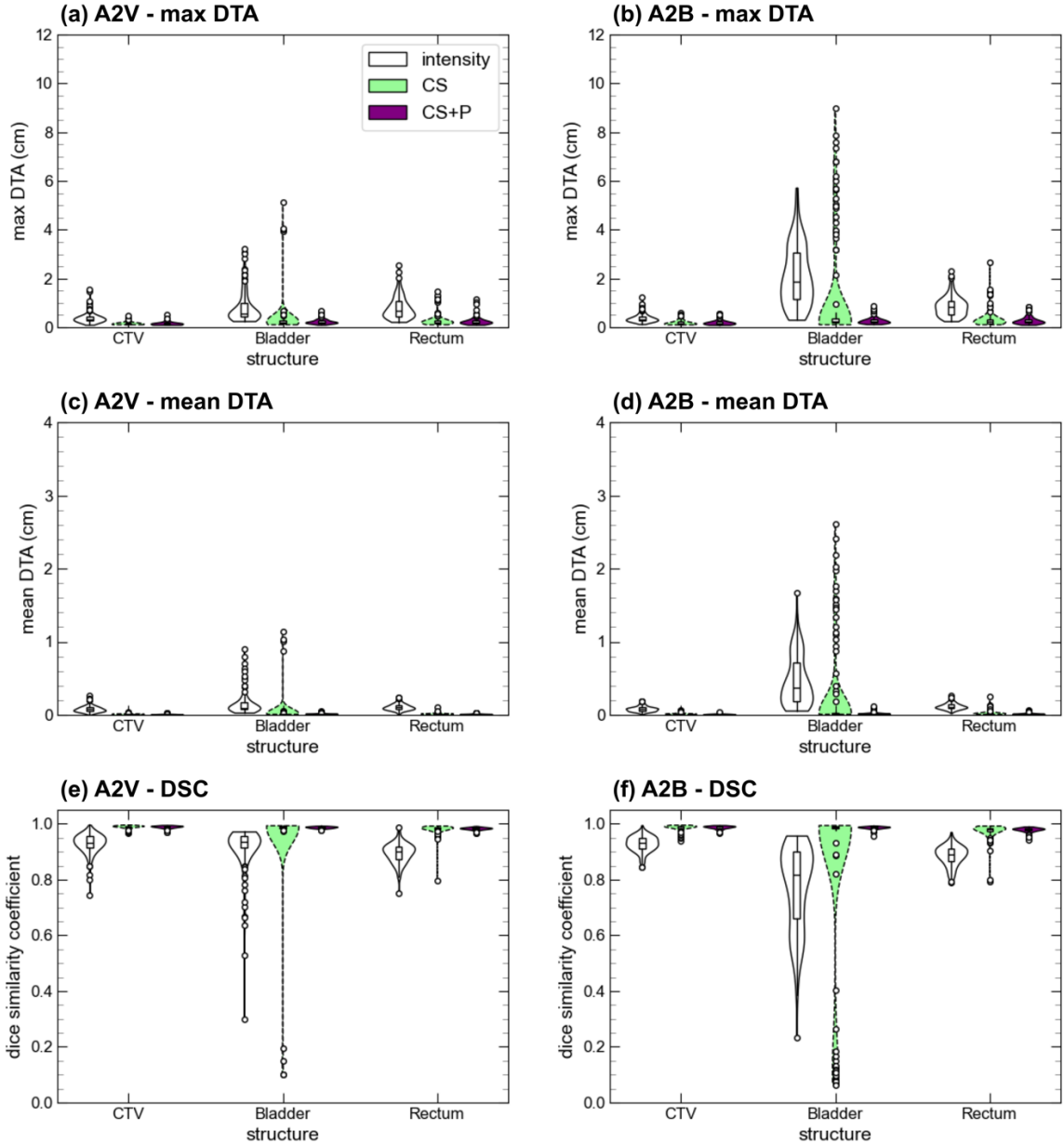

Figure 3: Intra-fractional geometry metrics comparing deformed CTV, bladder, rectum from the $MR_{adapt}$ to the $MR_{verify}$ (A2V) and $MR_{beam-on}$ (A2B) for the three DIR strategies (intensity-only, CS, CS+P).

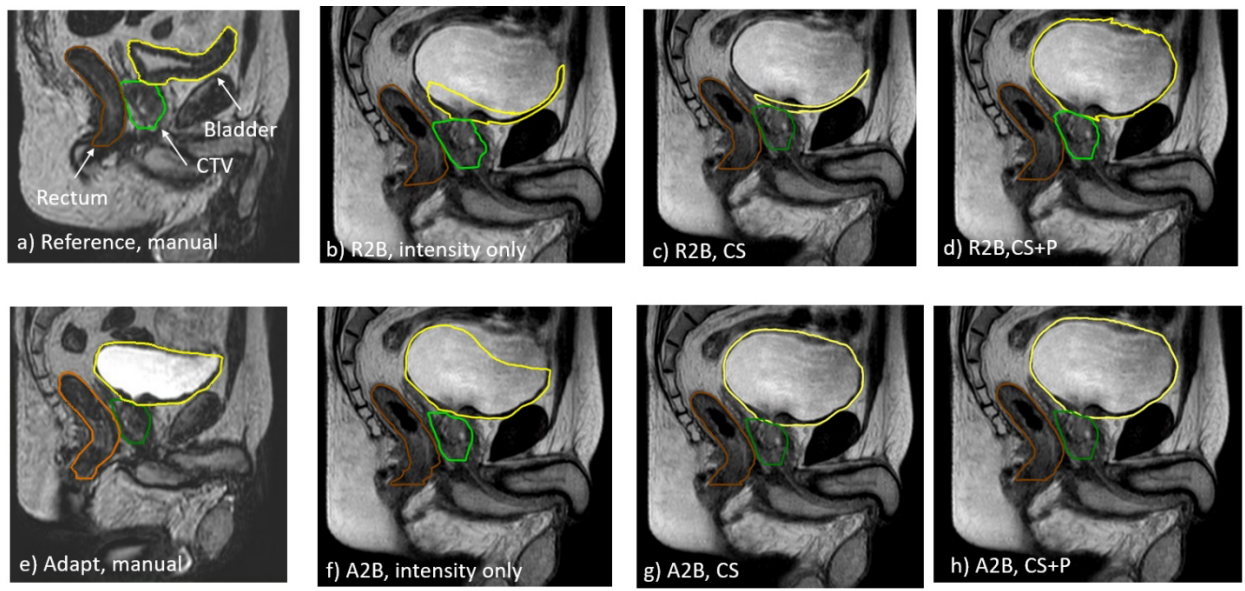

Figure 4: Comparison of different contour propagation methods to the beam-on image dataset. For both the reference (a) and adapt (e) images, contours are manually created. For beam on images (b-d, f-h), contours are propagated from either the reference (b-d) or adapt (f-h) image. Propagation based on either an intensity only (a,d), controlling ROI (b,e), or controlling ROI and POI (c,f) DIR approach are presented.

Table 1: Intra-fractional mean geometry metrics for the A2V and A2B image pairs for all three DIR strategies and contours. Values in brackets are $1\sigma$.

| image pair | DIR | OAR | DTA$_{max}$ (cm) | DTA$_{mean}$ (cm) | DSC |
|---|---|---|---|---|---|
| A2V | Intensity | CTV | 0.40 (0.05) | 0.083 (0.009) | 0.929 (0.007) |
| | | Bladder | 0.85 (0.13) | 0.15 (0.03) | 0.910 (0.018) |
| | | Rectum | 0.81 (0.09) | 0.110 (0.009) | 0.896 (0.007) |
| | CS | CTV | 0.153 (0.012) | 0.010 (0.001) | 0.989 (0.001) |
| | | Bladder | 0.36 (0.15) | 0.05 (0.04) | 0.96 (0.03) |
| | | Rectum | 0.28 (0.5) | 0.015 (0.002) | 0.982 (0.004) |
| | CS+P | CTV | 0.165 (0.014) | 0.010 (0.001) | 0.989 (0.001) |
| | | Bladder | 0.239 (0.020) | 0.018 (0.001) | 0.987 (0.001) |
| | | Rectum | 0.26 (0.03) | 0.014 (0.001) | 0.983 (0.001) |
| A2B | intensity | CTV | 0.38 (0.03) | 0.085 (0.007) | 0.928 (0.006) |
| | | Bladder | 2.11 (0.24) | 0.47 (0.07) | 0.77 (0.03) |
| | | Rectum | 0.88 (0.08) | 0.118 (0.008) | 0.886 (0.007) |
| | CS | CTV | 0.193 (0.022) | 0.015 (0.002) | 0.985 (0.002) |
| | | Bladder | 1.21 (0.43) | 0.27 (0.12) | 0.85 (0.06) |
| | | Rectum | 0.34 (0.07) | 0.024 (0.006) | 0.974 (0.005) |
| | CS+P | CTV | 0.20 (0.02) | 0.012 (0.001) | 0.988 (0.001) |
| | | Bladder | 0.272 (0.025) | 0.022 (0.003) | 0.987 (0.001) |

| | | | |
|---|---|---|---|
| Rectum | 0.29 (0.03) | 0.018 (0.002) | 0.979 (0.002) |

In Figure 5, we present the A2B ΔDVH curves for the bladder, CTV, and rectum for the three DIR strategies (the A2V plots are available in Supplement Figure S1). We report the DVH and per fraction difference DVH metrics for all image pairs, DIR strategies, and contours in Supplement Table S3. We observed improvement of the ΔDVH curves with increasing DIR complexity. Specifically for the A2B mapping, the 95% confidence interval (CI) of the CTV D98% is [-121 cGy, 192 cGy], [-44 cGy, 14 cGy], and [-17 cGy, 21 cGy], bladder D5cc is [-195 cGy, 121 cGy], [-38 cGy, 199 cGy], and [-73 cGy, 12 cGy], and of the rectum D1cc is [-350 cGy, 250 cGy], [-30 cGy, 70 cGy], and [-45 cGy, 56 cGy], reported for intensity-only, CS, and CS+P. Consistent with the geometry metrics, the CS approach show a subset of patients with large differences in bladder dose volume metrics (particularly between 20% and 60% of the bladder volume) due to poor bladder contour propagation. CTV ΔDVH variation above 90% relative volume reduces most between intensity and the CS strategy, demonstrated by the improvement of the 95% CI curve in Figure 5. Similarly, the rectum ΔDVH is most improved when using CS, compared to intensity only.

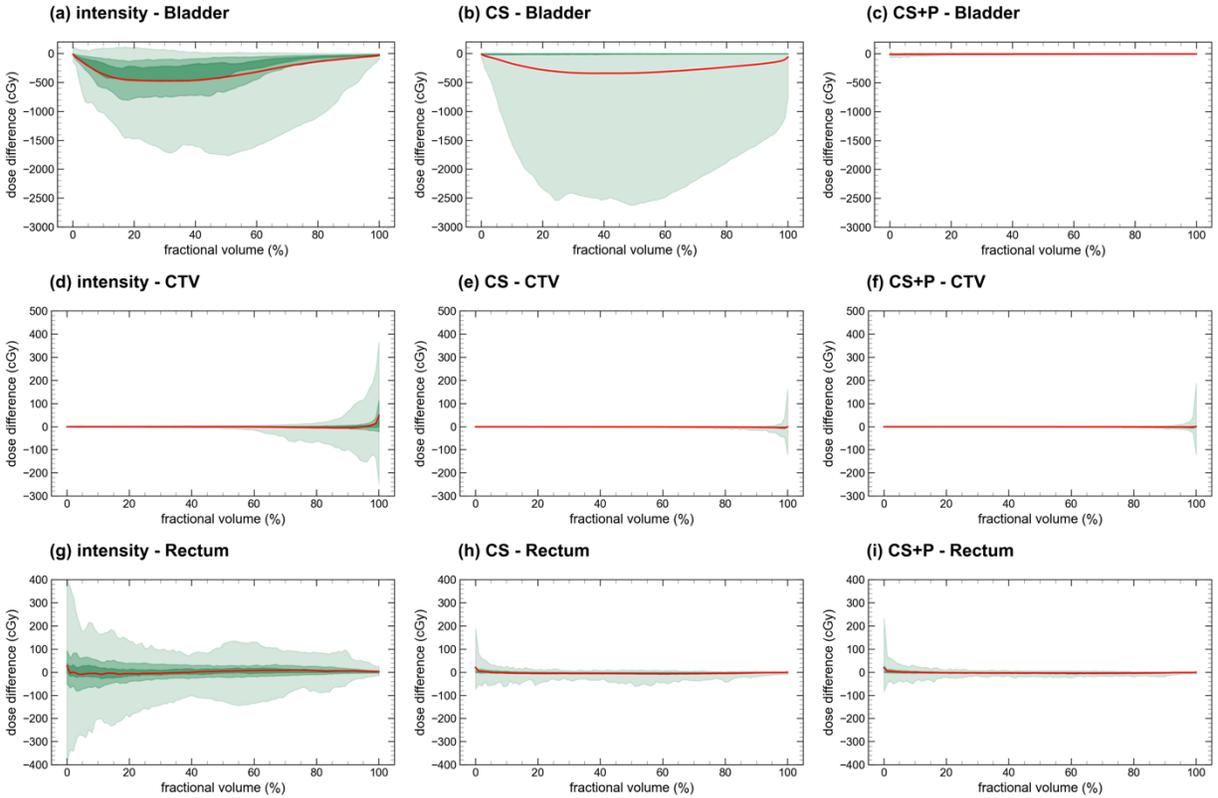

*Figure 5: adapt to beam-on (A2B) ΔDVH comparison for the three DIR strategies and contours. The green bands provide the 95% (lightest), 50% and 25% (darkest) confidence intervals, and the bold central red line represents the mean ΔDVH curve.*

**Inter-fractional**

In Figure 6, we present the geometry metrics for the R2A and R2B image pairs for all DIR strategies. Table 2 provides the metric means for the R2A, R2V, and R2B mappings. Similar to the intra-fractional results, we observe an improvement when using CS for the CTV and rectum. Comparing the intensity to CS DIRs, The $DTA_{mean}$ for the R2B mapping changed from 0.131 ± 0.009 cm to 0.018 ± 0.002 cm for the CTV and 0.154 ± 0.013 to 0.036 ± 0.013 cm for the rectum. However, for the bladder, though generally improved with the CS approach ($DTA_{mean}$ reducing from 0.46 ± 0.08 cm to 0.388 ± 0.14 cm for the R2B pair), a subset of fractions performs poorly with outliers most pronounced for the R2B pairing. In the supplement Figures 4S and 5S we present the DTA$_{max}$ as a function of relative bladder change for the A2B and R2B mappings for the three DIR strategies. For the intensity-only DIRs we see a consistent increase in DTA$_{max}$ as a function of bladder volume change. For the CS strategy, save for a few fractions, limited DTA$_{max}$ variation existed for < 100% bladder volume increase from reference, whereas for > 100% bladder volume increase (volume doubling) occurrence of DTA$_{max}$ > 2cm rises sharply. Applying the CS+P strategy corrects for these bladder deformation errors and reduces the magnitude of the DTA$_{max}$ as a function of bladder volume change (Figure S5c) as well as reducing DTA$_{means}$ to 0.015 ± 0.001 cm for the CTV, 0.025 ± 0.004 cm for the bladder, and 0.021 ± 0.002 cm for the rectum.

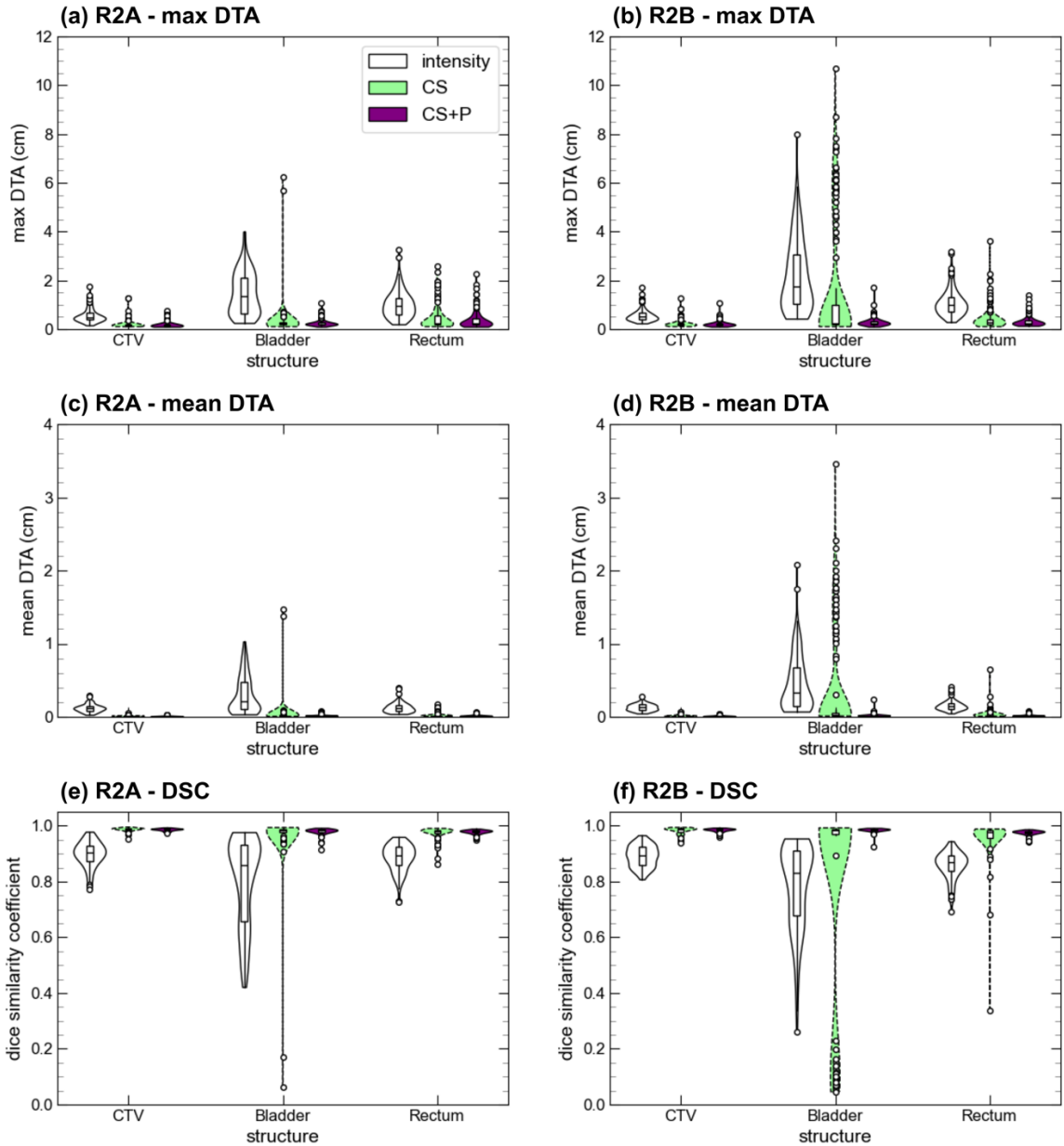

Figure 6: Inter-fractional geometry comparison for the R2A and R2B image pairs using the three DIR strategies for the CTV, bladder, and rectum.

*Table 2: Inter-fractional geometry comparison metrics of manual and deformed structures for all inter-fraction image combinations and DIR strategies. Values in brackets are 1$\sigma$.*

| image pair | DIR | OAR | DTA$_{max}$ (cm) | DTA$_{mean}$ (cm) | DSC |
|---|---|---|---|---|---|
| R2A | intensity | CTV | 0.57 (0.06) | 0.120 (0.010) | 0.899 (0.009) |
| | | Bladder | 1.48 (0.18) | 0.31 (0.05) | 0.80 (0.03) |
| | | Rectum | 1.02 (0.11) | 0.126 (0.013) | 0.886 (0.010) |
| | CS | CTV | 0.19 (0.03) | 0.012 (0.001) | 0.987 (0.001) |
| | | Bladder | 0.34 (0.15) | 0.04 (0.04) | 0.967 (0.022) |
| | | Rectum | 0.45 (0.10) | 0.022 (0.005) | 0.976 (0.004) |
| | CS+P | CTV | 0.185 (0.021) | 0.012 (0.001) | 0.987 (0.001) |
| | | Bladder | 0.259 (0.027) | 0.020 (0.002) | 0.981 (0.002) |
| | | Rectum | 0.39 (0.07) | 0.019 (0.002) | 0.978 (0.002) |
| R2V | intensity | CTV | 0.63 (0.06) | 0.143 (0.012) | 0.882 (0.010) |
| | | Bladder | 1.28 (0.17) | 0.23 (0.04) | 0.868 (0.021) |
| | | Rectum | 0.99 (0.09) | 0.149 (0.012) | 0.863 (0.010) |
| | CS | CTV | 0.20 (0.03) | 0.014 (0.002) | 0.985 (0.002) |
| | | Bladder | 0.64 (0.29) | 0.12 (0.07) | 0.93 (0.04) |
|   |   | Rectum | 0.37 (0.07) | 0.024 (0.006) | 0.972 (0.007) |
|   |   | CTV | 0.205 (0.027) | 0.013 (0.001) | 0.986 (0.001) |
|   | CS+P | Bladder | 0.272 (0.024) | 0.022 (0.002) | 0.985 (0.001) |
|   |   | Rectum | 0.31 (0.04) | 0.018 (0.002) | 0.979 (0.001) |
|   |   | CTV | 0.57 (0.05) | 0.131 (0.009) | 0.892 (0.008) |
|   | intensity | Bladder | 2.21 (0.29) | 0.46 (0.08) | 0.78 (0.03) |
|   |   | Rectum | 1.10 (0.11) | 0.154 (0.013) | 0.859 (0.010) |
|   |   | CTV | 0.22 (0.03) | 0.018 (0.002) | 0.982 (0.002) |
| R2B | CS | Bladder | 1.6 (0.5) | 0.388 (0.14) | 0.78 (0.08) |
|   |   | Rectum | 0.44 (0.10) | 0.036 (0.013) | 0.961 (0.013) |
|   |   | CTV | 0.222 (0.025) | 0.015 (0.001) | 0.985 (0.001) |
|   | CS+P | Bladder | 0.29 (0.04) | 0.025 (0.004) | 0.986 (0.001) |
|   |   | Rectum | 0.33 (0.05) | 0.021 (0.002) | 0.976 (0.002) |

In Figure 7, we present the R2B $\Delta$DVH curves for the bladder, CTV, and rectum for the three DIR strategies (R2A and R2V plots are detailed in Supplement Figures S2 and S3). The DVH and per fraction difference DVH metrics for all image pairs, DIR strategies, and contours are reported in Table S4. The $\Delta$DVH curves show improvement with increasing DIR complexity, consistent with the geometry metrics. Specifically, the 95% CI intensity-only, CS, and CS+P for the CTV D98% are [-126 cGy, 389 cGy], [-29 cGy, 19 cGy], and [-18 cGy, 26 cGy], for bladder D5cc are [-190 cGy, 245 cGy], [-51 cGy, 544 cGy], and [-79 cGy, 36 cGy], and for rectum D1cc are [-319 cGy, 247 cGy], [-106 cGy, 72 cGy], and [-52 cGy, 74 cGy]. Intensity-only distributions for the CTV and rectum are broader for the CTV and rectum compared to the intra-fraction results. Both CS and CS+P strategies produce comparable results to the intra-fractional data across all structures.

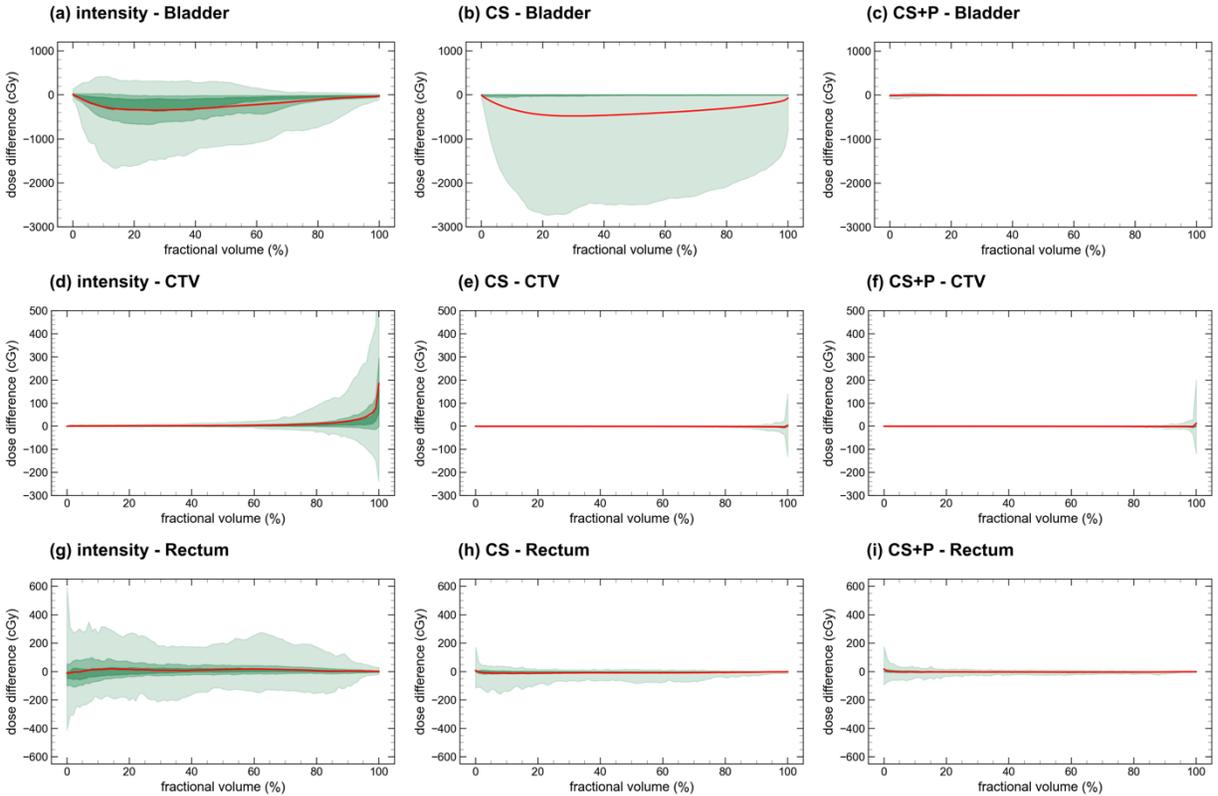

*Figure 7: Reference to beam-on (R2B) ΔDVH comparison for the three DIR strategies. The green bands provide the 95% (lightest), 50% and 25% (darkest) confidence intervals, and the bold central red line represents the mean ΔDVH curve.*

## Discussion

In this study, we performed a comprehensive analysis of multiple DIR strategies, including a novel hybrid structure and point DIR, applied to a cohort of patients with prostate cancer receiving MR-guided adaptive radiotherapy. Our overreaching goal was to demonstrate a general methodology for geometric and dosimetric evaluation of DIR accuracy and to arrive at a validated DIR strategy for dose accumulation in this patient cohort. A key strength of this work is the inclusion of multiple intra- and inter-fraction DIRs across 125 fractions with three different strategies for 1875 DIRs in total. Applying the DVFs to CTV, bladder, and rectum enabled comparison against expert manual contours using geometry and DVH metrics. Comparing DIR strategies, our novel controlling structure and point hybrid DIR strategy offered optimal DIR performance for both intra- and inter-fraction comparison.

In the context of clinical dose accumulation, inclusion of structures or points in the DIR must be justified based on the potential resource burden of segmenting key structures on verification or beam on images. Our results show that intensity-only DIR produced mean $DTA_{mean}$ values < 1.5 mm and DSC scores > 0.859 for the CTV and rectum structures for all image pairs. The bladder structure produced mean $DTA_{mean}$ values below 3.1 mm and DSC scores above 0.80 for the A2V, R2A, and R2V mappings while the mappings to the $MR_{beam-on}$ resulted in worse geometry metrics. TG-132 provides guidance on DIR evaluation and indicates a tolerance for $DTA_{mean}$ in the range of 2-3 mm and for DSC values above 0.8-0.9 (Brock et al., 2017). Our intensity-only DIRs for the CTV and rectum would meet these geometric criteria. However, when we incorporate DVH analysis for the intensity-only DIRs, we find that the different 95% CI for CTV D98% was as broad as [-126 cGy, 389 cGy], rectum D1cc was [-190 cGy, 245 cGy] and bladder D5cc was [-318 cGy, 247 cGy]. For the CTV, this D98% variation translates to a difference of -4.2 to 13.0% of the prescribed dose for our cohort. This demonstrates the importance of placing DIR validation in context of the delivered dose distribution and inspecting the entire population geometry metric distribution, as mean values may obscure individual patient/fraction deviations (Bohoudi et al., 2019; Vickress et al., 2017). Comparatively, Christiansen et al. looked at conventionally-fractionated prostate radiotherapy obtaining MR scans throughout their treatment course and, using the default Monaco DIR algorithm, obtained DSC values of 0.9, 0.87, 0.92 and $DTA_{mean}$ values of 1.0 mm, 1.25 mm, 1.11 mm for the prostate, rectum, and bladder, respectively, when comparing deformed and manual contours (Christiansen et al., 2020). These geometric results for the rectum and prostate are inline with our study. The bladder potentially improved compared to our intensity-only DIRs, however, these studies did not include a dosimetric evaluation to compare to our work. Previous reports have highlighted importance of DIR strategy evaluation beyond geometric analysis, including DVH analysis and visual inspection, as key DIR validation elements (Lowther et al., 2020; McDonald et al., 2023; Murr, Wegener, et al., 2024). Murr et. al. multi-institution comparison DSC values for the two prostate cases were 0.89 – 0.98 for CTV, 0.69 – 0.98 for rectum and 0.7- 0.99 for bladder, with the largest dosimetric differences observed for bladder V28Gy of 10.2% for prostate case 1 and 7.6% for prostate case 2.

To improve the DIR accuracy, we used the CTV, bladder, and rectum as input for the CS DIR strategy. We found this resulted in improved contour deformation for the CTV and rectum (mean $DTA_{mean}$ < 0.36 mm and DSC > 0.961 for all image pairs). The use of controlling structures has been shown to improve DIR mapping in abdominal and pelvic sites (Baroudi et al., 2023; McCulloch et al., 2022). Bohoudi et al. showed that CS DIR strategy for MR-to-MR mapping improved dose accumulation results for the rectum and bladder (Bohoudi et al., 2019) using film measurements on an anthropomorphic phantom. However, in our cohort, although there is a general improvement in the propagated bladder structures, a subset of fractions had substantial contour deviations resulting in a mean $DTA_{mean}$ as large as 3.9 mm and DSC as low as 0.78. Overall, we found that DIR performance was the worst for registrations with $MR_{beam-on}$, in which bladder volume differences between fixed and moving images were greatest; suggesting that bladder volume change is contributing to poor DIR performance for certain fractions. We inspected the relationship between the $DTA_{max}$ and relative bladder volume change and observed for the intensity-only DIRs that $DTA_{max}$ consistently increased with increasing change in bladder volume (Supplemental Figure S4). For the CS strategy, a sharp decrease in DIR performance is observed when the bladder doubles in size across image pairs. However, many DIRs performed well for the large bladder deformations beyond this threshold. This suggests that there is a component of patient specific bladder anatomy impacting the CS DIR performance. Adaptive prostate treatment sessions can exceed 50 minutes, over which continued bladder filling occurs (Li et al., 2023). Patients in our study received ART-specific bladder preparation instructions and with rigorous instruction and management instances of restarting adaptive sessions reduced, but did not significantly change the initial bladder volumes (Dang et al., 2022). The impact of more stringent bladder filling may translate into improved CS DIR performance but as noted above, the bladder shape may be the substantive contributing factor to poor DIRs.

The RayStation ANACONDA algorithm uses a chamfer matching distance metric for the CS algorithm which has been observed to produce erroneous DIRs in the case of large deformations (Weistrand & Svensson, 2015b). Starting with the RayStation 2023B version, an additional image similarity metric has been added to help mitigate this issue and would require further validation for this application (Lorenzo Polo et al., 2024). In the absence of this updated algorithm, we propose a novel controlling point generation and produce guiding points on the surfaces of the rectum, CTV, and bladder structures to drive the CS+P DIR strategy. We show this DIR approach corrects for the bladder deformation errors, DSC above 0.97 and $DTA_{mean}$ below 0.21 mm for all image pairs, while maintaining performance for the prostate and rectum. Comparing the R2B CS and CS+P DIRs, we see a change in the CTV D98% from [-29 cGy, 19 cGy] to [-18 cGy, 26 cGy], rectum D1cc from [-106 cGy, 72 cGy] to [-52 cGy, 74 cGy], and bladder D5cc from [-51 cGy, 544 cGy] to [-79 cGy, 36 cGy]. Measures of D98% results are within 1-2% of the clinical prescription/goal when using the CS+P approach. Based on the improved geometry and DVH analysis of the CS+P DIR, we selected this strategy as the approach for future dose accumulation work on this cohort.

Though the CS+P strategy is generalizable to other sites and image modality pairs, DIR evaluation per site, dose distribution, and relevant clinical DVH metrics would be necessary. Not all disease sites will be impacted by systematic organ changes over the adaptive treatment course, as in pelvic sites, and may face other challenges such as the influence of peristaltic motion. For liver and upper gastro-intestinal targets, intensity-only or CS approaches may yield acceptable results when confined to an area of interest for DVH analysis (McCulloch et al., 2022; Semeniuk et al., 2024). Incorporation of controlling points or use of the updated ANACONDA algorithm for these sites could be an interest for future exploration. Controlling structure approaches hinge on the availability of contours across paired images to drive the controlling point generation and the DIR mapping. This introduces observer contouring variability and currently limits automation of the CS+P DIR strategy. For expert contour delineation, MR-based inter-observer variability was evaluated by Pathmanathan et al. for prostate with a DSC of 0.94 and $DTA_{max}$ of 4.8 mm (Pathmanathan et al., 2019), and Sanders et al. for the prostate, rectum, and bladder with DSC of 0.904, 0.893, and 0.928, respectively (Sanders et al., 2022). Christiansen et al. evaluated MR-based intra-observer variability with

DSC of 0.92, 0.95, and 0.97, $DTA_{mean}$ of 0.88 mm, 0.65mm, and 0.55mm, and $DTA_{max}$ of 4.89 mm, 7.65 mm, and 4.05 mm, for the prostate, rectum and bladder (Christiansen et al., 2020).

Automated contouring for both conventional and adaptive treatment planning is seeing a substantial uptake in the field. As McCulloch et al. proposed for liver patients, we expect that using auto-contours to establish structures for the CS/CS+P approaches will reduce or eliminate manual intervention (McCulloch et al., 2022). Though requiring further validation, if using auto contouring specifically to drive DIR mapping, we hypothesize that even if differences between automated and manual contours exist, so long as the automated contours are anatomically consistent for a given patient across fractions and within-session images the resulting CS/CS+P DIRs would likely be acceptable. The use of sufficiently reliable automated contouring would address the limitations of the CS/CS+P approach and allow the process to be run passively on a per-patient basis with minimal user inspection or intervention.

## Conclusions

In this study, we have rigorously analyzed DIR strategies for MR guided radiotherapy prostate patients, focusing on the evaluation intensity-only, controlling structure (CS), and controlling structure with points (CS+P). We demonstrated improved performance of the CS+P strategy, which better accounts for bladder filling during the adaptive fraction. We incorporate and highlight the importance of including dose metrics in DIR validation where use of only geometry metrics may obscure DIR inaccuracies. Our dosimetric analysis shows a 1-2% uncertainty (95% CI) in clinically relevant dose metrics for the CS+P DIR strategy. The selected methodology can be used for dose accumulation studies of MRL prostate patients and applicability may be expanded by introduction of auto contouring.

# Supplemental Material

*Table S3: Institutional clinical goals for 3000 cGy in 5 fractions prostate SBRT*

| Structure | Metric | Criteria |
|---|---|---|
| CTVp_3000 | D95 | > 3300cGy |
| PTVp_3000 | D0cc | < 4000cGy |
|  | D2 | < 3500cGy |
|  | D95 | > 3000cGy |
|  | D98 | > 2850cGy |
| Rectum | D50 | < 1000cGy |
|  | D20 | < 2000 cGy |
|  | D1cc | < 3000 cGy |
| Bladder | D40 | < 1500 cGy |
|  | D5cc | < 3000 cGy |
| Femur_L/R | D5 | < 1200 cGy |
| SmallBowel | D1cc | < 2500 cGy |
| LargeBowel | D1cc | < 2500 cGy |
| PenileBulb | D50 | < 2400 cGy |
| PenileBulb | D1cc | < 3000 cGy |
| Urethra | D50 | < 3500 cGy |

*Table S4: MR Sequence parameters*

| Parameter | $T_2$-weighted 6-minute scan | $T_2$-weighted 2-minute scan |
|---|---|---|
| Field-of-view (mm) | 400 x 400 x 250 | 400 x 400 x 300 |
| Echo time (ms) | 82 | 278 |
| Repetition time (ms) | 1300 | 1535 |
| SENSE | 4 (phase), 1.2 (slice) | 3.6 (phase) |
| Half-scan | 0.6 | 0.625 |
| Acquisition resolution (mm) | 1.2 x 1.2x1.2 | 1.2 x 1.2x1.2 |
| Reconstructed resolution (mm) | 0.5 x 0.5 x 0.6 | 0.5 x 0.5 x 1.0 |
| Acquisition time | 6 min 7 s | 1 min 57 s |

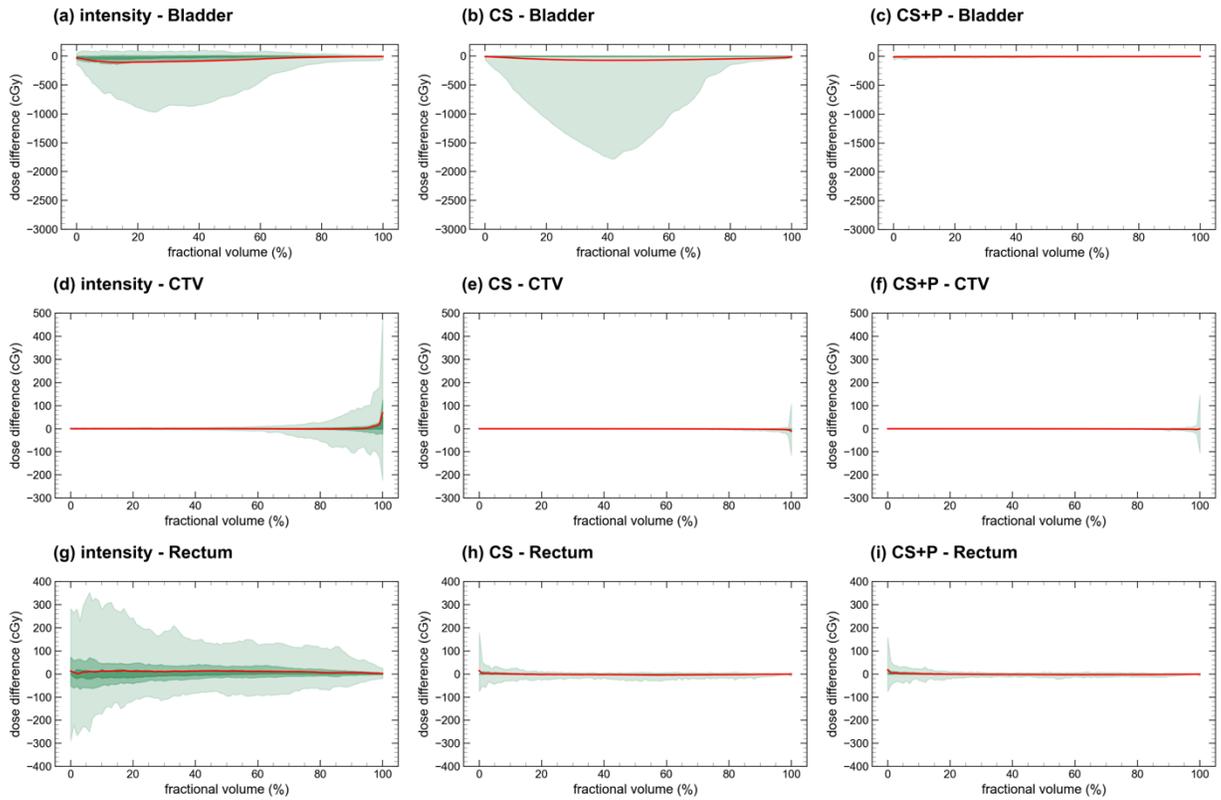

*Figure S8: adapt to verify (A2V) ΔDVH comparison for the three DIR strategies. The green bands provide the 95% (lightest), 50% and 25% (darkest) confidence intervals, and the bold central red line represents the mean ΔDVH curve.*

*Table 5S: DVH metrics and per fraction DVH metric difference means for intra-fraction image combinations for the intensity, CS, CS+P DIR strategies. Values in brackets are 1σ.*

| image pair | DIR | OAR | Metric | Manual mean (cGy) | Mapped mean (cGy) | Mean Manual – Mapped (cGy) |
|---|---|---|---|---|---|---|
| A2V | intensity | CTV | D98% | 3150 (270) | 3140 (280) | 10 (110) |
|  |  | Bladder | D5cc | 2990 (230) | 3030 (220) | -43 (93) |
|  |  | Rectum | D1cc | 2240 (660) | 2250 (650) | -0 (140) |
|  | CS | CTV | D98% | 3150 (270) | 3160 (270) | -3.7 (6.3) |
|  |  | Bladder | D5cc | 2990 (230) | 2990 (240) | 1 (41) |
|  |  | Rectum | D1cc | 2240 (660) | 2240 (670) | 7 (17) |
|  | CS+P | CTV | D98% | 3150 (270) | 3150 (270) | -2.2 (7.0) |
|  |  | Bladder | D5cc | 2990 (230) | 2990 (230) | -6 (12) |
|  |  | Rectum | D1cc | 2240 (660) | 2230 (670) | 7 (18) |
| A2B | intensity | CTV | D98% | 3120 (280) | 3120 (270) | 8 (75) |
|  |  | Bladder | D5cc | 3010 (270) | 3030 (270) | -21 (88) |
|  |  | Rectum | D1cc | 2220 (650) | 2220 (650) | 0.0 (150) |

|     |         |      |            |            |           |
| --- | ------- | ---- | ---------- | ---------- | --------- |
|     | CTV     | D98% | 3120 (280) | 3130 (280) | -5.0 (14) |
| CS  | Bladder | D5cc | 3010 (270) | 3000 (290) | 15 (91)   |
|     | Rectum  | D1cc | 2220 (650) | 2210 (650) | 11 (26)   |
|     | CTV     | D98% | 3120 (280) | 3130 (280) | -1 (11)   |
| CS+P| Bladder | D5cc | 3010 (270) | 3020 (260) | -8 (22)   |
|     | Rectum  | D1cc | 2220 (650) | 2210 (650) | 9 (27)    |

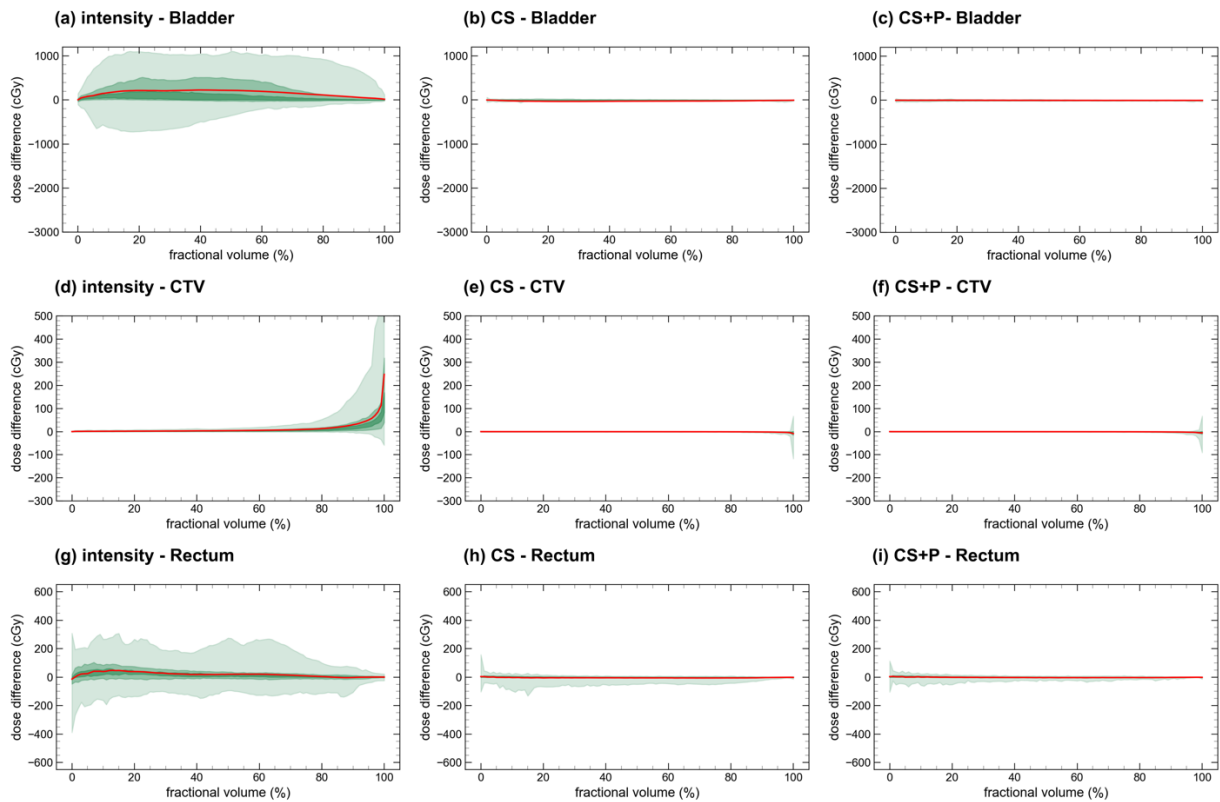

Figure S9: Reference to adapt (R2A) $\Delta$DVH comparison for the three DIR strategies. The green bands provide the 95% (lightest), 50% and 25% (darkest) confidence intervals, and the bold central red line represents the mean $\Delta$DVH curve.

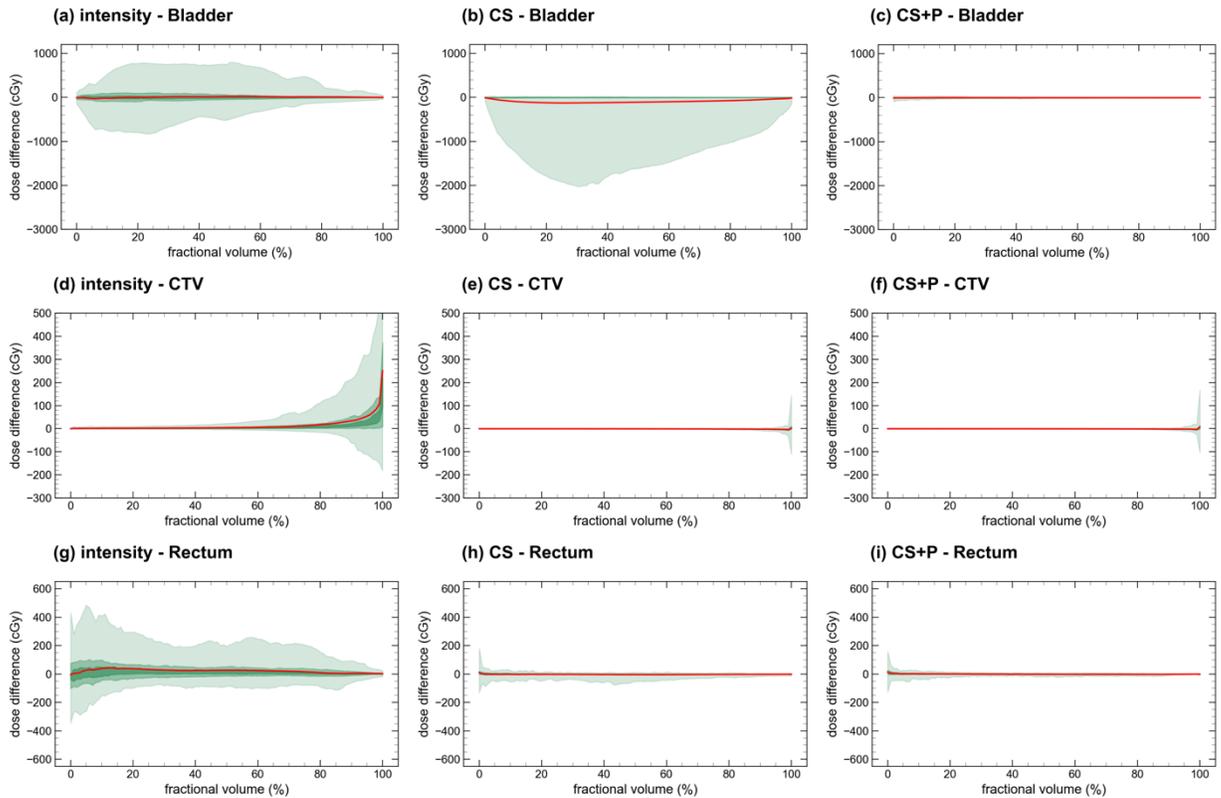

*Figure S10: Reference to verify (R2V) ΔDVH comparison for the three DIR strategies. The green bands provide the 95% (lightest), 50% and 25% (darkest) confidence intervals, and the bold central red line represents the mean ΔDVH curve.*

*Table S6: DVH metrics and per fraction DVH metric difference means for inter-fraction image combinations for the intensity, CS, CS+P DIR strategies. Values in brackets are 1σ.*

| image pair | DIR | OAR | Metric | Manual mean (cGy) | Mapped mean (cGy) | Mean Manual – Mapped (cGy) |
|---|---|---|---|---|---|---|
| R2A | intensity | CTV | D98% | 3265 (32) | 3180 (150) | 90 (140) |
|  |  | Bladder | D5cc | 2890 (190) | 2850 (240) | 40 (160) |
|  |  | Rectum | D1cc | 2530 (570) | 2540 (560) | -10 (100) |
|  | CS | CTV | D98% | 3265 (32) | 3269 (31) | -3.5 (4.7) |
|  |  | Bladder | D5cc | 2890 (190) | 2880 (190) | 10 (53) |
|  |  | Rectum | D1cc | 2530 (570) | 2520 (570) | 5 (24) |
|  | CS+P | CTV | D98% | 3265 (32) | 3269 (31) | -3.6 (5.8) |
|  |  | Bladder | D5cc | 2890 (190) | 2890 (180) | 2 (16) |
|  |  | Rectum | D1cc | 2530 (570) | 2520 (580) | 5 (24) |
| R2V | intensity | CTV | D98% | 3150 (270) | 3070 (340) | 90 (220) |
|  |  | Bladder | D5cc | 2990 (230) | 3000 (240) | -10 (130) |

|     |           | Rectum  | D1cc  | 2240 (660) | 2260 (650) | -20 (160) |
|-----|-----------|---------|-------|------------|------------|-----------|
|     |           | CTV     | D98%  | 3150 (270) | 3160 (270) | -4 (11)   |
|     | CS        | Bladder | D5cc  | 2990 (230) | 2970 (260) | 12 (90)   |
|     |           | Rectum  | D1cc  | 2240 (660) | 2240 (660) | 4 (25)    |
|     |           | CTV     | D98%  | 3150 (270) | 3150 (270) | -2.3 (8.0)|
|     | CS+P      | Bladder | D5cc  | 2990 (230) | 2990 (220) | -6 (25)   |
|     |           | Rectum  | D1cc  | 2240 (660) | 2240 (670) | 6 (21)    |
|     |           | CTV     | D98%  | 3120 (280) | 3070 (300) | 60 (130)  |
|     | intensity | Bladder | D5cc  | 3010 (270) | 3010 (290) | 10 (110)  |
|     |           | Rectum  | D1cc  | 2220 (650) | 2240 (640) | -20 (160) |
|     |           | CTV     | D98%  | 3120 (280) | 3130 (280) | -3 (15)   |
| R2B | CS        | Bladder | D5cc  | 3010 (270) | 2970 (310) | 40 (130)  |
|     |           | Rectum  | D1cc  | 2220 (650) | 2220 (650) | -1 (43)   |
|     |           | CTV     | D98%  | 3120 (280) | 3130 (280) | -1 (14)   |
|     | CS+P      | Bladder | D5cc  | 3010 (270) | 3020 (260) | -6 (25)   |
|     |           | Rectum  | D1cc  | 2220 (650) | 2220 (650) | 5 (33)    |

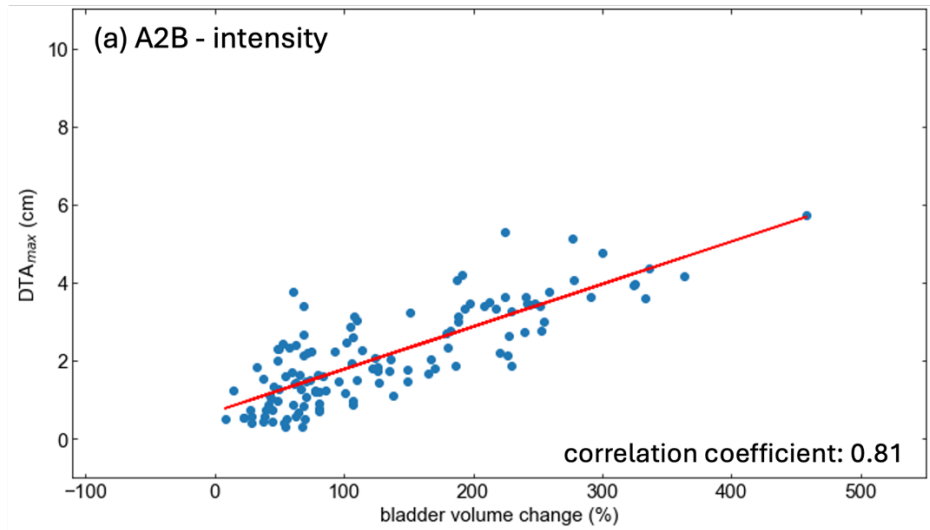
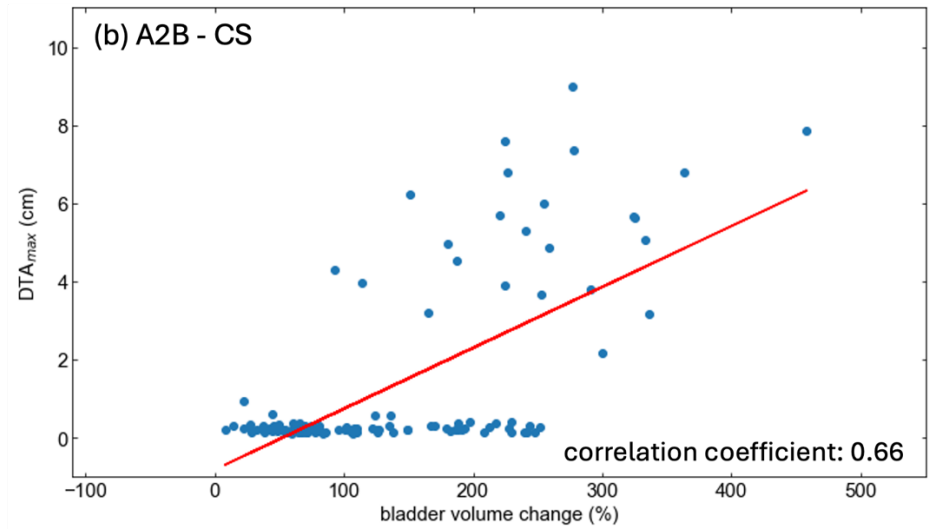
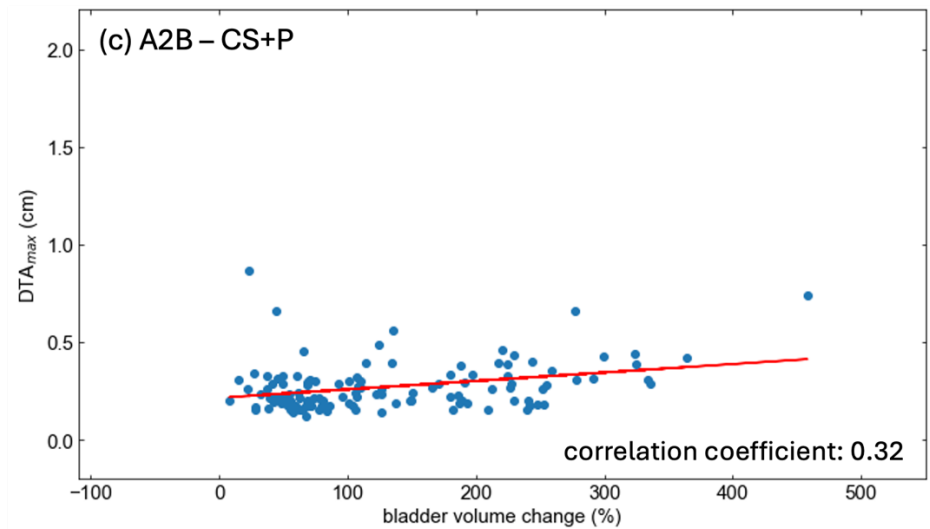

Figure S11: DTA$_{max}$ as a function of bladder volume change between the MR$_{adapt}$ and the MR$_{beam-on}$ for the (a) intensity (b) CS and (c) CS+P DIR strategies. Note the change of scale in (c).

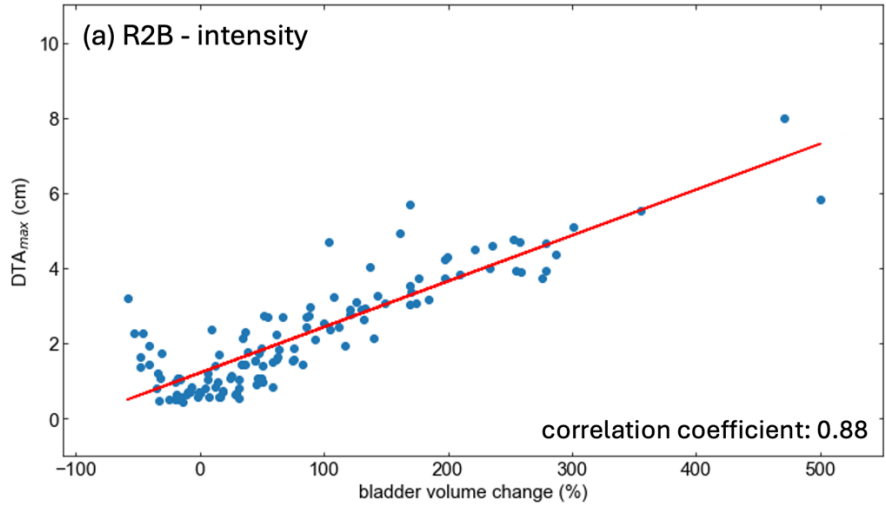
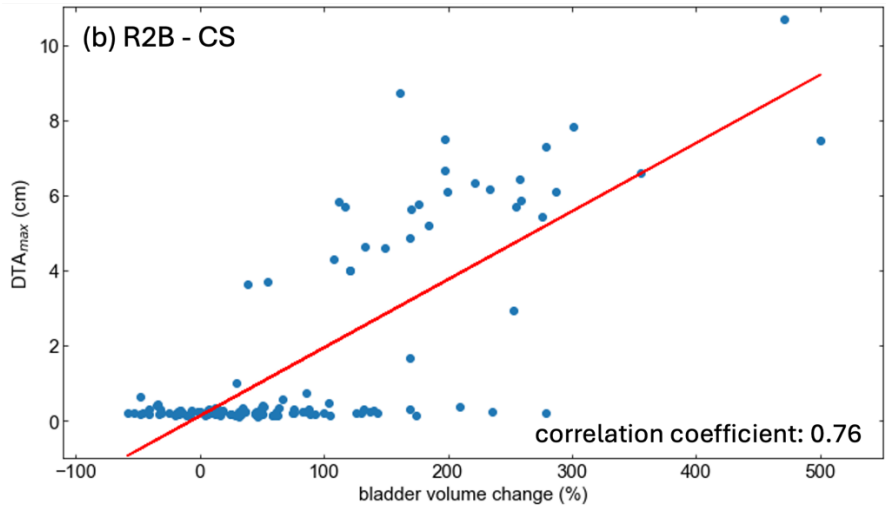
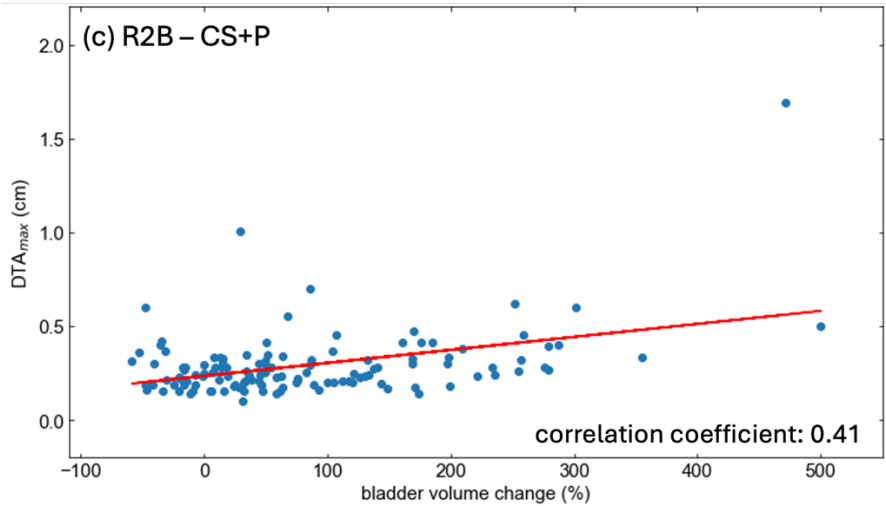

*Figure S12: DTA$_{max}$ as a function of bladder volume change between the MR$_{ref}$ and the MR$_{beam-on}$ for the (a) intensity (b) CS and (c) CS+P DIR strategies. Note the change of scale in (c)*